\documentclass[useAMS,usenatbib]{mn2e} \usepackage{epsfig}
\usepackage{amsmath}
\usepackage{natbib}

\author[Stephan and Fortenberry] {Cody J. Stephan and Ryan C.
Fortenberry\thanks{Email:rfortenberry@georgiasouthern.edu}  \\
Georgia Southern University, Department of Chemistry \& Biochemistry, Statesboro,
GA 30460 U.S.A.}

\title[HeHHe$^+$, HeHNe$^+$, \& HeHAr$^+$]{The Interstellar Formation and
Spectra of the Noble Gas, Proton-Bound HeHHe$^+$, HeHNe$^+$, \& HeHAr$^+$
Complexes}

\begin{document}

\date{Submitted: \today}

\maketitle

\begin{abstract}

The sheer interstellar abundance of helium makes any bound molecules or
complexes containing it of potential interest for astrophysical observation.
This work utilizes high-level and trusted quantum chemical techniques to
predict the rotational, vibrational, and rovibrational traits of HeHHe$^+$,
HeHNe$^+$, and HeHAr$^+$.  The first two are shown to be strongly bound, while
HeHAr$^+$ is shown to be more of a van der Waals complex of argonium with a
helium atom.  In any case, the formation of HeHHe$^+$ through reactions of
HeH$^+$ with HeH$_3$$^+$ is exothermic.  HeHHe$^+$ exhibits the
quintessentially bright proton-shuttle motion present in all proton-bound
complexes in the 7.4 micron range making it a possible target for telescopic
observation at the mid-IR/far-IR crossover point and a possible tracer for the
as-of-yet unobserved helium hydride cation.  Furthermore, a similar mode in
HeHNe$^+$ can be observed to the blue of this close to 6.9 microns.  The
brightest mode of HeHAr$^+$ is dimmed due the reduced interaction of the helium
atom with the central proton, but this fundamental frequency can be found
slightly to the red of the Ar$-$H stretch in the astrophysically detected
argonium cation.

\begin{keywords}
astrochemistry -- molecular data -- molecular processes -- ISM: molecules --
radio lines: ISM -- infrared: ISM
\end{keywords}

\end{abstract}

\section{Introduction}

Helium and hydrogen make up nearly all of the observable matter in the universe
leaving chemists to squabble over the remaining scraps.  These scraps are what
compose the planets, our bodies, and most other things engineered by human
beings.  Nearly all other processes depend upon atoms much more interesting
than the first two on the periodic table.  Even so, helium and hydrogen can
engage in chemistry with one another almost certainly combining to make HeH$^+$
\citep{Hogness25}.  This cation should be produced in detectable amounts if for
no other reason than the sheer abundance of the constituents in the
interstellar medium (ISM) \citep{Roberge82}.  However, such an interstellar
observation of this diatomic cation has yet to be reported in the literature.
It was the analogous ArH$^+$ that has been observed toward various astronomical
sources\citep{Barlow13, Schilke14, Roueff14, Neufeld16} making the argonium and
not helonium (helium hydride) cation the first noble gas molecule detected in
nature.  The smaller and more abundant helium and even neon hydride cations
have not been observed, yet.

The chemistry of helium is likely the least voluminous for any of the elements
between hydrogen and iron even in controlled laboratory conditions.  However,
helium will make complexes and form some bonds.  Helium cationic clusters have
been predicted, He$_m$H$_n$$^+$ clusters have been synthesized, dication
complexes observed, and even hydrogen-like replacement structures analyzed
\citep{Frenking90, Roth01, Grandinetti04, Savic15, Zicler16}.  In all cases,
the issue is that the helium cation binding in any of these complexes is
relatively weak making long-lifetime molecules and high enough abundances for
observable interstellar spectra of such chemical combinations quite unlikely.

Like unto helium, neon is reluctant to form bonds.  There is little surprise
here due to the high ionization potentials and relatively poor polarizabilities
in these smallest of noble gas compounds \citep{Taylor89, Rice91, Pauzat05,
Pauzat07, Pauzat09, Pauzat13}.  Neonium (NeH$^+$) has been well-characterized
\citep{Ram85, Matsashima98, Gamallo13, Koner16, Coxon16}, but it has yet to be
conclusively observed in any astrophysical environment.  While the reaction of
Ar$^+$ with ubiquitous hydrogen gas leads to ArH$^+$ and hydrogen atoms in the
ISM, the analogous reaction with neon will initially lead to neutral neon atoms
and ionized hydrogen gas \citep{RTheis15}.  More complicated neon structures
beyond NeH$^+$ have been proposed and even synthesized, but few have bond
strengths in the covalent range \citep{Frenking90, Grandinetti11}.  Notable
exceptions include NeOH$^+$ and NeCCH$^+$ recently characterized theoretically
at high level \citep{RTheis16, Novak17}, but these are several factors less
stable than their argon counterparts.  With helium and neon being so very
abundant in the ISM, neon even more so than nitrogen \citep{Savage96},
molecules containing these atoms may still be awaiting detection.

These molecules in waiting could be proton-bound complexes.  These structures
involve a mostly bare proton situated between two other atoms or molecules
where the mutual attraction of the ligands to the proton creates fairly strong
interactions.  These structures are of additional significance to
astrochemistry and astrophysical observation due to their proton ``rattle'' or
``shuttle'' motion.  In such vibrational modes, basically only the proton
moves.  Hence, little of the mass but nearly all of the charge is moving
creating an immense change in dipole moment.  As a result, such vibrational
transitions are incredibly strong absorbers/emitters meaning that small column
densities of materials are required to create observable spectral features.
OCHCO$^+$, NNHNN$^+$, and the heteromolecular combinations have been analyzed
recently showing that these bright vibrational modes can be found from the
near- to mid- and even far-IR wavelengths \citep{Terrill10, Cotton12,
Fortenberry15OCHCO+, Yu15NNHNN+, Fortenberry16NNHCO+, Fortenberry16COHNN+,
Begum16} making them tantalizing targets for the epoch for growth in IR
telescopic power in which we are currently in the midst.

Proton-bound complexes of noble gases have been known for some time.  In fact,
the simplest, HeHHe$^+$, was noted for its relatively strong bonds nearly 35
years ago \citep{Dykstra82}.  However, a complete and reliable set of
rovibrational spectroscopic data have yet to be produced for this simple system
while other insights into its nature have been explored theoretically
\citep{Baccarelli97, Panda03, Bartl13}.  Other data for related noble gas
molecules have been produced including those with helium \citep{Fridgen98,
Lundell99, Koner12, Koner14, Borocci15, Grabowski16, Koner16}, but full
spectral charactization is still lacking for most of these structures.

Very recently, the vibrational spectra of Ar$_n$H$^+$ complexes were
characterized experimentally including ArHAr$^+$ \citep{McDonald16} with its
bright proton motion at 10.11 microns with dissociation not occurring until 1.74
microns (or 0.711 eV).  Simultaneously, theoretical work on this complex
produced a very similar dissociation energy (0.719 eV) and comparable
vibrational frequencies \citep{Fortenberry17NgHNg+}.  The NeHNe$^+$ and
NeHAr$^+$ complexes were also analyzed.  \cite{Fortenberry17NgHNg+} showed that
the NeHNe$^+$ complex is actually more strongly bound than ArHAr$^+$ indicating
that, for once, the neon bonds are actually stronger than the more polarizable
argon bonds in a cation.  The NeHNe$^+$ dissociation is higher at 0.867 eV, but
interstellar synthesis of NeHNe$^+$ is most likely in the gas phase from
reactions of NeH$^+$ with NeH$_3$$^+$.  Again, the former has yet to be
observed, and the latter is only weakly bound \citep{RTheis152}.  ArHAr$^+$ is
also favorably created from ArH$^+$ and ArH$_3$$^+$ where the former is, again,
known in the ISM and the latter is a viable interstellar candidate
\citep{Pauzat05, Pauzat13, RTheis152}.  Additionally, ArHAr$^+$ has a much
brighter and longer wavelength proton shuttle motion making it more likely to
be observed in the ISM \citep{Fortenberry17NgHNg+}.

Consequently, the question lingers as to whether proton-bound complexes
involving the abundant helium atom are viable interstellar detection
candidates.  Furthermore, combinations of noble gas atoms in such complexes
with helium are known to be fairly stable \citep{Koner12, Grabowski16} and
those with other noble gas atoms have been classified at high-level with good
comparison to experiment \citep{Fortenberry17NgHNg+}.  As a result, this work
will employ the same methodology as that utilized previously on proton-bound
complexes \citep{Fortenberry16COHNN+, Fortenberry16NNHCO+, Fortenberry17NgHNg+}
where comparison in other molecules to gas phase experimental results has
provided exceptional accuracy on the order of 0.01 micron accuracy for
vibrational features and 30 MHz for rotational constants \citep{Huang08,
Huang09, Huang11, Zhao14, Fortenberry11HOCO, Fortenberry11cHOCO,
Fortenberry12HOCO+, Huang13NNOH+, Huang13C3H+, Fortenberry13C3H-,
Fortenberry14C2H3+, Fortenberry15SiC2, Kitchens16, Fortenberry16C2H3+}.  These
data will be useful for the spectral characterization of such molecules in the
ISM with current and upcoming ground- and space-based telescopes such as the
upcoming James Webb Space Telescope.  Potentially expanding the noble gas
molecular budget of the ISM will grow our understanding of interstellar
chemical bonding and provide novel chemical pathways for these so-called
``inert'' and noble gases.

\section{Computational Details}

The $ab\ initio$, quantum chemical computational methodology employed here has
been detailed elsewhere \citep{Huang08, Huang09, Huang11, Fortenberry11HOCO}
and specifically for proton-bound complexes of noble gas cation dimers by
\cite{Fortenberry17NgHNg+}.  For completeness, coupled cluster theory
\citep{ccreview, Shavitt09} at the singles, doubles, and perturbative triples
[CCSD(T)] level \citep{Rag89} is employed in all computations within the PSI4
quantum chemistry package \citep{psi4}.  The geometries of these linear
complexes are treated at the aug-cc-pV5Z level \citep{Dunning89, aug-cc-pVXZ,
cc-pVXZ} and corrected for core correlation with the Martin-Taylor (MT) core
correlating basis set \citep{Martin94}.  From these geometries, 0.005 \AA\
displacements of the bond lengths and 0.005 radian displacements of the bond
angle within the symmetry-internal coordinates defined below are computed to
produce a fourth-order Taylor series expansion of the internuclear Hamiltonian
called a quartic force field (QFF).  The coordinates for HeHHe$^+$ are defined
as:
\begin{align}
S_1(\Sigma_g)&=\frac{1}{\sqrt{2}}[(\mathrm{He}_1-\mathrm{H})+(\mathrm{He}_2-\mathrm{H})]\\
S_2(\Sigma_u)&=\frac{1}{\sqrt{2}}[(\mathrm{He}_1-\mathrm{H})-(\mathrm{He}_2-\mathrm{H})]\\
S_3(\Pi_{xz})&=\angle\mathrm{He}-\mathrm{H}-\mathrm{He}-{\bf{y}}\\
S_4(\Pi_{yz})&=\angle\mathrm{He}-\mathrm{H}-\mathrm{He}-{\bf{x}}.
\end{align}
Those for the HeHNe$^+$ and HeHAr$^+$ molecules are produced from (with Ng
representing either Ne or Ar):
\begin{align}
S_1(\Sigma^+) &= \mathrm{He}-\mathrm{H}\\
S_2(\Sigma^+) &= \mathrm{Ng}-\mathrm{H}\\
S_3(\Pi_{xz}) &= \angle\mathrm{He}-\mathrm{H}-\mathrm{Ng}-{\bf{y}}\\
S_4(\Pi_{yz}) &= \angle\mathrm{He}-\mathrm{H}-\mathrm{Ng}-{\bf{x}}.
\end{align}
As a consequence of the differences in geometrical connectivities and atomic
symmetries, HeHHe$^+$ requires the use of 57 total points, while HeHNe$^+$ and
HeHAr$^+$ require 69 to define the QFF.

At each displacement point, CCSD(T)/aug-cc-pVTZ, -pVQZ, and -pV5Z energies are
extrapolated to the complete basis set (CBS) limit via a three-point formula
\citep{Martin96} and augmented for inclusion of core correlation from the MT
basis sets as well as scalar relativity \citep{Douglas74}.  This produces the
CcCR QFF for ``CBS,'' ``core correlation,'' and ``relativity.'' The
relativistic computations are made with the MOLPRO 2010.1 quantum chemistry
package \citep{molpro}.  A least-squares fitting of the points produces the
equilibrium geometry, and refitting the points produces zero gradients and the
subsequent force constants.  The fitting is tight with a sum of squared
residuals on the order of 10$^{-17}$ a.u.$^2$ for the Ne and Ar complexes and
10$^{-18}$ a.u.$^2$ for HeHHe$^+$.  The force constants are transformed from
the symmetry-internal coordinates into more generic Cartesian coordinates with
the INTDER program \citep{intder}.  Then, second-order vibrational
perturbation theory (VPT2) and rotational perturbation theory are employed to
provide the frequencies and spectroscopic constants \citep{Mills72, Watson77,
Papousek82} within the SPECTRO program \citep{spectro91}.  Double-harmonic
intensities are computed with the Gaussian09 program and the MP2/6-31+G$^*$
level of theory \citep{MP2, Hehre72, g09} which has been shown to provide
notable agreement for the vibrational intensities of NNHNN$^+$ with a more
advanced and time-consuming semi-global dipole moment surface
\citep{Yu15NNHNN+}.

\begingroup
\begin{table}
\caption{CCSD(T)/aug-cc-pV5Z Formation/Destruction Energetics.}
\label{RelEnergy}
\centering
\small
\begin{tabular}{l | c c}
\hline
Reactions & kcal/mol & eV \\
\hline
He + HeH$^+$ $\rightarrow$ HeHHe$^+$ & -13.2 & -0.57 \\
2He + H$^+$ $\rightarrow$ HeHHe$^+$ & -60.3 & -2.62 \\
2HeH$^+$ $\rightarrow$ HeHHe$^+$ + H$^+$ & 33.8 & 1.47 \\
HeH$^+$ + HeH$_3$$^+$ $\rightarrow$ HeHHe$^+$ + H$_3$$^+$ & -11.9 & -0.52\\
He + H$_3$$^+$ $\rightarrow$ HeH$_3$$^+$ & -1.3 & -0.06\\
\hline
He + NeH$^+$ $\rightarrow$ HeHNe$^+$ & -11.9 & -0.52 \\
Ne + HeH$^+$ $\rightarrow$ HeHNe$^+$ & -21.9 & -0.95 \\
HeH$^+$ + NeH$^+$ $\rightarrow$ HeHNe$^+$ + H$^+$ & 35.2 & 1.52 \\
NeH$^+$ + HeH$_3$$^+$ $\rightarrow$ HeHNe$^+$ + H$_3$$^+$ & -10.6 & -0.46\\
HeH$^+$ + NeH$_3$$^+$ $\rightarrow$ HeHNe$^+$ + H$_3$$^+$ & -15.4 & -0.67\\
Ne + H$_3$$^+$ $\rightarrow$ NeH$_3$$^+$ & -6.5 & -0.28\\
\hline
He + ArH$^+$ $\rightarrow$ HeHAr$^+$ & -2.1 & -0.09 \\
Ar + HeH$^+$ $\rightarrow$ HeHAr$^+$ & -49.0 & -2.13 \\
HeH$^+$ + ArH$^+$ $\rightarrow$ HeHAr$^+$ + H$^+$ & 45.0 & 1.95 \\
ArH$^+$ + HeH$_3$$^+$ $\rightarrow$ HeHAr$^+$ + H$_3$$^+$ & -0.8 & -0.03\\
HeH$^+$ + ArH$_3$$^+$ $\rightarrow$ HeHAr$^+$ + H$_3$$^+$ & -40.0 & -1.73\\
Ar + H$_3$$^+$ $\rightarrow$ ArH$_3$$^+$ & -9.1 & -0.39\\
\hline

\end{tabular}
\end{table}
\endgroup

\section{Results and Discussion}

The interstellar presence for any of these proton-bound complexes can be
established by observing the necessary rotational or vibrational transitions.
However, the possible formation of these species must be established in order
for such searches to be deemed even plausible before observation can begin.  In
light of such, high-level computations at the CCSD(T)/aug-cc-pV5Z level are
performed here in order to determine if these proton-bound complexes can form
exothermically in the gas-phase.  Similar and previous analysis
\citep{Fortenberry17NgHNg+} shows that ArHAr$^+$ is a likely interstellar
proton-bound complex created from the reaction of ArH$^+$ with ArH$_3$$^+$.
Table \ref{RelEnergy} contains these data for HeHHe$^+$, HeHNe$^+$, and
HeHAr$^+$ in the standard chemistry unit of kcal/mol as well as in eV.  The
complexes lie to the right of the reaction arrow indicating that negative
energies are desirously exothermic in the gas phase.

HeHHe$^+$ is thermodynamically stable since removal of a single helium atom has
a 0.57 eV energy cost.  Removal of both helium atoms is over five times higher
in energy.  The collision of two helium hydride cations will not form HeHHe$^+$
and a hydrogen atom.  However and like with ArHAr$^+$, the reaction of helium
hydride with helium trihydride produces HeHHe$^+$ with an excess of 0.52 eV of
energy.  The formation of HeH$_3$$^+$ is a matter of further discussion, but
HeH$_3$$^+$ is stable requiring a small but non-negligible amount of energy to
dissociate the helium atom.

\begingroup
\begin{table}

\caption{The HeHHe$^+$ CcCR Symmetry-Internal Force Constants (in
mdyn/\AA$^n$$\cdot$rad$^m$)$^a$.}

\label{HeHHefc}

\centering

\begin{tabular}{c r c r c r}
\hline

F$_{11}$ & 3.061 326 & F$_{331}$ & -0.3869 & F$_{2222}$ & 55.55 \\ 
F$_{22}$ & 0.632 926 & F$_{441}$ & -0.3869 & F$_{3322}$ & -0.46 \\ 
F$_{33}$ & 0.101 838 & F$_{1111}$ & 83.54 &  F$_{3333}$ &  0.37 \\ 
F$_{44}$ & 0.101 838 & F$_{2211}$ & 45.26 &  F$_{4422}$ & -0.46 \\ 
F$_{111}$ &-17.6653 &  F$_{3311}$ &  1.43 &  F$_{4433}$ &  0.26 \\ 
F$_{221}$ & -8.4929 &  F$_{4411}$ &  1.43 &  F$_{4444}$ &  0.37 \\ 
 
\hline
\end{tabular}

$^a$1 mdyn $=$ $10^{-8}$ N; $n$ and $m$ are exponents corresponding to the
number of units from the type of modes present in the specific force constant.
 
\end{table}
\endgroup
 
HeHNe$^+$ is also thermodynamically stable, but removal of the helium atom
(-0.52 eV) is nearly half as energetically costly as removal of the neon atom
at -0.95 eV.  This is likely actually the strongest neon bond produced thus far
besides neonium itself.  The neon bonding in NeOH$^+$ is -0.53 eV, NeCCH$^+$ is
-0.93 eV, and NeHNe$^+$ -0.87 eV.  Helonium reacting with NeH$_3$$^+$ favors
creation of HeHNe$^+$ and the ubiquitous H$_3$$^+$ with -0.67 eV.  Reacting
neonium with He$_3$$^+$ is also exothermic but less favorable at -0.46 eV.

\begingroup
\begin{table*}

\caption{The CcCR Zero-Point ($R_{\alpha}$ vibrationally-averaged) and
Equilibrium Structures, Rotation Constants, Vibrationally Excited-Rotation
Constants, Distortion Constants, and Vibrational Frequencies and
Intensities$^a$ of HeHHe$^+$ and Its Various Isotopologues.}

\label{HeHHe}

\centering

\small

\begin{tabular}{l | l l l l l l} 
\hline\hline

                   & HeHHe$^+$ & HeDHe$^+$ & $^{3}$HeHHe$^+$ & $^{3}$HeDHe$^+$ & $^{3}$HeH$^{3}$He$^+$ & $^{3}$HeD$^{3}$He$^+$ \\
\hline
r$_0$(He$-$H) \AA  & 0.946 482 & 0.942 915 & 0.949 044 & 0.945 724 & 0.947 882 & 0.944 400 \\
r$_e$(He$-$H) \AA  & 0.924 908 & 0.924 908 & 0.924 908 & 0.924 908 & 0.924 908 & 0.924 908 \\
$B_e$ MHz          & 73798.7 & 73798.7 & 85651.9 & 85484.6 & 97938.9 & 97938.9 \\
$B_0$ MHz          & 70816.5 & 71217.8 & 82106.4 & 82598.6 & 81486.9 & 82214.5 \\
$B_1$ MHz          & 64474.8 & 67405.1 & 74980.1 & 80428.6 & 73398.9 & 79997.2 \\
$B_2$ MHz          & 71730.3 & 71673.0 & 83067.1 & 82814.7 & 82474.5 & 82322.6 \\
$B_3$ MHz          & 69366.2 & 69929.3 & 80220.4 & 78564.1 & 79269.6 & 77340.9 \\
$D_e$ MHz          &   1.378 &   1.378 &  1.860  &   1.858 &   2.428 &   2.428 \\
$H_e$ $\mu$Hz      &  12.491 &  12.491 &  18.765 &  16.940 &  29.196 &  29.196 \\
\hline                                                    
$\omega_1(\sigma_u)$ Proton Shuttle cm$^{-1}$ & 1549.3 (2661) & 1155.5 & 1566.5 & 1249.6 & 1577.3 & 1312.5 \\
$\omega_2(\pi)$ Bend cm$^{-1}$                &  950.2 (294)  &  708.7 &  958.9 &  720.2 &  967.4 &  731.6 \\
$\omega_3(\sigma_g)$ Symm.~He$-$H cm$^{-1}$   & 1139.4        & 1139.4 & 1225.0 & 1152.4 & 1312.5 & 1192.9 \\
$\nu_1(\sigma_u)$ Proton Shuttle cm$^{-1}$    & 1345.2        & 1030.8 & 1347.1 &  987.3 & 1357.2 &  955.0 \\
$\nu_2(\pi)$ Bend cm$^{-1}$                   &  884.9        &  670.2 &  891.5 &  680.1 &  898.1 &  689.8 \\
$\nu_3(\sigma_g)$ Symm.~He$-$H cm$^{-1}$      &  895.9        &  917.1 &  956.3 & 1016.1 & 1007.5 & 1055.6 \\
$2\nu_1$ cm$^{-1}$                            & 2870.6        & 2194.0 & 2864.5 & 1734.4 & 2920.4 & 1950.6 \\
$2\nu_2$ cm$^{-1}$                            & 1869.6        & 1416.9 & 1889.2 & 1456.1 & 1908.6 & 1501.3 \\
$2\nu_3$ cm$^{-1}$                            & 1760.7        & 1804.5 & 1868.1 & 1925.6 & 1975.6 & 2275.7 \\
$\nu_1+\nu_2$ cm$^{-1}$                       & 2003.3        & 1577.4 & 2008.3 & 1567.6 & 2021.8 & 1578.2 \\
$\nu_1+\nu_3$ cm$^{-1}$                       & 1925.3        & 1681.3 & 1985.0 & 2158.7 & 1979.1 & 1707.6 \\
$\nu_2+\nu_3$ cm$^{-1}$                       & 1723.2        & 1528.1 & 1782.9 & 1588.9 & 1832.8 & 1615.0 \\
ZPE cm$^{-1}$                                 & 2266.6        & 1830.1 & 2322.0 & 1891.7 & 2375.7 & 1951.9 \\
\hline

\hline
\end{tabular}
\\ $^a$MP2/6-31+G$^*$ double-harmonic vibrational intensities in km/mol in
parentheses.  The modes where the intensities are zero by symmetry have no
values.
\end{table*}

\endgroup

HeHAr$^+$ is much more dichotomous.  The helium is not well-bound (-0.09 eV),
but the argon atom very much is (-2.13 eV).  This trend is carried through the
other formation and destruction reactions listed in Table \ref{RelEnergy}.
However, ArH$_3$$^+$ has the strongest Ng$-$H bond of all the noble
gas-trihydride cations and is hypothesized to exist in the ISM
\citep{RTheis152, Pauzat13}.  Reacting this specie with the almost guaranteed
interstellar HeH$^+$ produces HeHAr$^+$ and -1.73 eV of energy.

Hence, HeHAr$^+$ is the most energetically-favored product, but interstellar
abundances still point to HeHHe$^+$ as the most likely to be observed.  In any
case, only spectroscopic observation in the ISM can determine the presence for
any of these compounds.  The following data for each complex should be able to
assist in the laboratory or even interstellar observation of these noble gas,
molecular cations.

\subsection{HeHHe$^+$}

The force constants, those needed to construct the QFF potential within the
Hamiltonian, for the proton-bound, helium cation dimer are given in Table
\ref{HeHHefc}.  The harmonic, diagonal force constants can be viewed as
proportional to bond strength.  Breaking these symmetry-internal coordinates
(from Eqs. 1-4) down into simple-internal coordinates, i.e. He$-$H coordinates,
produces a 2.165 mdyne/\AA$^2$ force constant for the He$-$H stretch.  This is
is nearly identical to the 2.158 mydne/\AA$^2$ Ne$-$H force constant in
NeHNe$^+$ and greater than that in ArHAr$^+$ \citep{Fortenberry17NgHNg+}.  This
is greater even than the force constant in NeOH$^+$ \citep{RTheis16}.
Consequently, the notable bond energy discussed previously for the removal of
the helium atom from HeHHe$^+$ is corroborated by the internal molecular
structure, as well.

Further corroboration comes from the He$-$H bond length given in Table
\ref{HeHHe}.  This value is 0.946 \AA, which is 0.17 \AA\ longer than the same
bond in the helium hydride cation.  However, the He$-$H bond length is still
significantly less than bond lengths in helium van der Waals complexes which
often approach 3 \AA.  The rotational constants are also given even though
HeHHe$^+$ possesses no permanent dipole moment.  The distortion constants are
fairly large since the molecule has a fairly small mass and a labile proton.
On the other hand, the vibrationally-excited rotational constants ($B_{\nu}$),
notably $B_1$ and $B_2$, can be utilized for rovibrational spectral modeling.

\begingroup
\begin{table}

\caption{The HeHNe$^+$ CcCR Reduced Simple-Internal Force Constants (in
mdyn/\AA$^n$$\cdot$rad$^m$).}

\label{HeHNefc}

\centering

\begin{tabular}{c r c r c r}
\hline

F$_{11}$ & 1.560 258 & F$_{331}$ & -0.2807 & F$_{3311}$ &  0.58 \\
F$_{21}$ & 1.169 047 & F$_{332}$ & -0.2682 & F$_{3321}$ &  1.02 \\
F$_{22}$ & 2.174 774 & F$_{441}$ & -0.2807 & F$_{3322}$ &  0.56 \\
F$_{33}$ & 0.098 183 & F$_{442}$ & -0.2682 & F$_{3333}$ &  0.46 \\
F$_{44}$ & 0.098 183 & F$_{1111}$ &  81.86 & F$_{4411}$ &  0.58 \\
F$_{111}$ & -12.8685 & F$_{2111}$ &   8.08 & F$_{4421}$ &  1.02 \\
F$_{211}$ &  -3.2887 & F$_{2211}$ &  12.05 & F$_{4422}$ &  0.56 \\
F$_{221}$ &  -2.9577 & F$_{2221}$ &   6.37 & F$_{4433}$ &  0.29 \\
F$_{222}$ & -16.9096 & F$_{2222}$ & 115.52 & F$_{4444}$ &  0.46 \\                      
\hline
\end{tabular}


\end{table}
\endgroup

\begingroup
\begin{table*}

\caption{The CcCR HeHNe$^+$ Zero-Point ($R_{\alpha}$ vibrationally-averaged)
and Equilibrium Structures, Rotation Constants, Vibrationally Excited-Rotation
Constants, Distortion Constants, Vibrational Frequencies, and Intensities
(km/mol).}

\label{HeHNe}

\centering

\small

\begin{tabular}{l | l l l l l l l l}
\hline\hline

		   & HeHNe$^+$ & HeDNe$^+$ & HeH$^{22}$Ne$^+$ &
HeD$^{22}$Ne$^+$ & $^{3}$HeHNe$^+$ & $^{3}$HeDNe$^+$ & $^{3}$HeH$^{22}$Ne$^+$ &
$^{3}$HeD$^{22}$Ne$^+$ \\

\hline
r$_0$(He$-$H) \AA  & 0.977 376 & 0.975 914 & 0.977 417 & 0.975 988 & 0.980 477 & 0.979 002 & 0.980 526 & 0.979 079 \\
r$_0$(Ne$-$H) \AA  & 1.124 207 & 1.119 893 & 1.124 079 & 1.119 722 & 1.123 164 & 1.118 759 & 1.123 035 & 1.118 588 \\
r$_e$(He$-$H) \AA  & 0.953 323 & 0.953 323 & 0.953 323  & 0.953 323 & 0.953 323 & 0.953 323 & 0.953 323 & 0.953 323 \\
r$_e$(Ne$-$H) \AA  & 1.113 207 & 1.113 207 & 1.113 207  & 1.113 207 & 1.113 207 & 1.113 207 & 1.113 207 & 1.113 207 \\
$B_e$ MHz          & 34117.0 & 32946.6 & 33525.5 & 32307.7 & 42551.9 & 40411.7 & 41940.7 & 39737.2 \\
$B_0$ MHz          & 33114.8 & 32187.5 & 32543.3 & 31567.2 & 41259.1 & 39464.1 & 40669.8 & 38810.7 \\
$B_1$ MHz          & 30502.4 & 30432.1 & 29969.5 & 29841.4 & 37988.1 & 37398.9 & 37439.1 & 36783.9 \\
$B_2$ MHz          & 33671.5 & 32535.5 & 33087.3 & 31911.0 & 41957.4 & 39867.9 & 41360.4 & 39210.9 \\
$B_3$ MHz          & 32609.5 & 31728.8 & 32064.6 & 31124.2 & 40547.9 & 38826.5 & 39977.4 & 38184.1 \\
$D_e$ kHz          & 238.640 & 218.636 & 230.300 & 210.162 & 367.623 & 328.492 & 356.984 & 317.700 \\
$H_e$ mHz          &  86.700 & 323.489 &  92.429 & 308.120 & 481.355 & 536.845 & 473.786 & 496.507 \\
$\mu$ D            &  3.10   & \\
\hline                                                    
$\omega_1(\sigma)$ Shuttle cm$^{-1}$& 1569.8 (2610) & 1138.9 & 1568.3 & 1137.5 & 1577.3 & 1160.5 & 1576.0 & 1559.9 \\
$\omega_2(\pi)$ Bend cm$^{-1}$      &  824.3 (287)  &  605.2 &  824.0 &  604.7 &  833.4 &  617.5 &  833.0 &  617.0 \\
$\omega_3(\sigma)$ He$-$H cm$^{-1}$ &  856.9 (15)   &  852.2 &  850.0 &  844.3 &  963.0 &  945.1 &  956.4 &  936.9 \\
$\nu_1(\sigma)$ Shuttle cm$^{-1}$   & 1453.6        & 1035.7 & 1448.8 & 1033.6 & 1374.3 & 1019.9 & 1373.5 & 1017.4 \\
$\nu_2(\pi)$ Bend cm$^{-1}$         &  792.9        & 1257.0 &  792.6 &  584.4 &  798.8 &  594.6 &  798.5 &  594.2 \\
$\nu_3(\sigma)$ He$-$H cm$^{-1}$    &  691.2        &  714.9 &  686.0 &  710.0 &  766.5 &  801.4 &  762.1 &  797.3 \\
$2\nu_1$ cm$^{-1}$                  & 2961.0        & 2155.8 & 2957.7 & 2148.6 & 2958.5 & 2090.2 & 2954.4 & 2081.6 \\
$2\nu_2$ cm$^{-1}$                  & 1708.5        & 1242.2 & 1706.8 & 1240.3 & 1720.3 & 1271.9 & 1718.9 & 1269.8 \\
$2\nu_3$ cm$^{-1}$                  & 1303.4        & 1412.0 & 1298.2 & 1403.2 & 1532.5 & 1580.2 & 1524.3 & 1571.7 \\
$\nu_1+\nu_2$ cm$^{-1}$             & 2036.3        & 1523.9 & 2033.8 & 1519.2 & 2027.0 & 1503.6 & 2024.3 & 1500.8 \\
$\nu_1+\nu_3$ cm$^{-1}$             & 1959.6        & 1605.3 & 1951.8 & 1598.2 & 1979.3 & 1683.4 & 1972.6 & 1684.1 \\
$\nu_2+\nu_3$ cm$^{-1}$             & 1432.1        & 1257.0 & 1427.4 & 1252.9 & 1505.8 & 1347.5 & 1502.0 & 1344.5 \\
ZPE cm$^{-1}$                       & 2038.5        & 1590.3 & 2034.0 & 1585.2 & 2097.3 & 1654.2 & 2093.1 & 1649.4 \\
\hline                                                    

\hline
\end{tabular}
\end{table*}

\endgroup

Furthermore, the vibrational intensities indicate that the proton-shuttle
motion will be have a large transition moment/be a strong absorber or emitter
as would be expected for such a complex.  The double-harmonic 2661 km/mol
intensity for $\omega_1$ is nearly the same as NeHNe$^+$ and roughly half that
of ArHAr$^+$ which are all two orders of magnitude greater than most
vibrational intensities.  The anharmonic shuttle motion, $\nu_1$, lies at
1345.2 cm$^{-1}$ or 7.43 microns at the red end of the mid-IR.  Deuteration
drops this value into the far-IR at 1030.8 cm$^{-1}$ or 9.70 microns.
Inclusion of the lighter $^{3}$He isotope blueshifts the frequencies slightly
for each inclusion.  The $\nu_2$ bending mode also has a notable intensity and
is actually quite bright relative to more traditional vibrational modes.  The
anharmonic frequency for the bend (or perpendicular proton motion) is 884.9
cm$^{-1}$ or 11.30 microns.  The two-quanta combination bands and overtones
are also given in Table \ref{HeHHe}.  Those modes containing $\nu_1$ will also
be bright as has been shown for ArHAr$^+$ \citep{McDonald16}.

\subsection{HeHNe$^+$}

The HeHNe$^+$ force constants are given in Table \ref{HeHNefc}.  Immediately,
the He$-$H F$_{11}$ force constant shows weakening in this bond upon inclusion
of the neon atom.  However, the F$_{22}$ Ne$-$H force constant is quite close
to that in NeHNe$^+$ and actually is even greater by a small margin
corroborating the energetic data from Table \ref{RelEnergy}.  The He$-$H bond
length consequently grows to 0.977 \AA\ in HeHNe$^+$ relative to the helium
dimer, and the Ne$-$H bond length is small at 1.124 \AA, as shown in Table
\ref{HeHNe}.  This is shorter than the 1.156 \AA\ Ne$-$H bond length in
NeHNe$^+$ \citep{Fortenberry17NgHNg+}.

HeHNe$^+$ is rotationally active and has a large dipole moment of 3.10 D
computed with the center-of-mass at the origin from CCSD/aug-cc-pVTZ. The
dipole moment and subsequent rotational constants provided in Table \ref{HeHNe}
will assist in pointing closely towards the placement for the rotational lines
of this linear complex.  Isotopic substitution with $^{3}$He actually has more
of an effect than D in this case since the helium atoms are on the outside of
the molecule and the hydrogen is closer to the center-of-mass.  Again, the
vibrationally-excited rotational lines are also given in Table \ref{HeHNe} in
order to provide complete rovibrational analysis of HeHNe$^+$.  The magnitudes
of the distortion constants are not as large in this complex as they are in
HeHHe$^+$, and scale, at least for $D_e$, with the mass of neon relative to
helium.

The $\omega_1$ proton-shuttle motion in HeHNe$^+$ has a similar intensity at
2610 km/mol (Table \ref{HeHNe}) as HeHHe$^+$ and NeHNe$^+$.  The $\omega_3$
stretch can be classified as both the Ne and He atoms moving away from the
proton, but the lighter He atom has a larger vector leading to the qualitative
description of this state as the He$-$H stretch.  This fundamental will also be
visible, but the change in dipole moment is quite small creating only a 15
km/mol intensity.  The bend intensity is also in the same range as the bending
frequency intensities in HeHHe$^+$ and NeHNe$^+$.

Replacement with the neon atom to create HeHNe$^+$ actually blue-shifts the
bright $\nu_1$ fundamental vibrational frequency to 1453.6 cm$^{-1}$ (6.88
microns) relative to the bright mode in HeHHe$^+$.  The other two modes
red-shift as one would expect for vibrational frequencies involving a heavier
atom, neon in this case.  The reason for this likely lies in the non-zero, by
symmetry, F$_{21}$ mixed harmonic force constant that is forbidden in
HeHHe$^+$.  This additional overlap is quite large at 1.169 mdyne/\AA$^2$ from
Table \ref{HeHNefc} and allows the two stretches to couple further since more
symmetry-allowed avenues are opened.  Consequently, the total interatomic
interaction increases in the proton-sharing for HeHNe$^+$.  The $\nu_3$ stretch
in HeHNe$^+$ is further red-shifted relative to the helium proton-bound dimer
since the He$-$H F$_{11}$ force constant is computed here to be significantly
reduced in the CcCR QFF VPT2 computations.

Deuteration decreases the frequencies notably, but replacement with $^{3}$He
affects the He$-$H $\nu_3$ stretch more than it does the $\nu_2$ bend with
shifts of 106.1 cm$^{-1}$ and 9.1 cm$^{-1}$, respectively.  Inclusion of
$^{22}$Ne affects the frequencies, as well, but the shift in relative mass is
less obtrusive for this atom making its isotopic shifts far less than those for
D and $^3$He.

\subsection{HeHAr$^+$}

\begingroup
\begin{table}

\caption{The HeHAr$^+$ CcCR Reduced Simple-Internal Force Constants (in
mdyn/\AA$^n$$\cdot$rad$^m$).}

\label{HeHArfc}

\centering

\begin{tabular}{c r c r c r}
\hline

F$_{11}$ & 0.105 357 & F$_{331}$ & -0.0957 & F$_{3311}$ &  1.00 \\
F$_{21}$ & 0.193 842 & F$_{332}$ & -0.0510 & F$_{3321}$ & -0.35 \\
F$_{22}$ & 4.011 600 & F$_{441}$ & -0.0957 & F$_{3322}$ & -0.88 \\
F$_{33}$ & 0.027 336 & F$_{442}$ & -0.0510 & F$_{3333}$ &  0.61 \\
F$_{44}$ & 0.027 336 & F$_{1111}$ &  3.07 &  F$_{4411}$ &  1.00 \\
F$_{111}$ & -0.6979 &  F$_{2111}$ &  2.12 &  F$_{4421}$ & -0.35 \\
F$_{211}$ & -0.7749 &  F$_{2211}$ &  1.21 &  F$_{4422}$ & -0.88 \\
F$_{221}$ & -0.2108 &  F$_{2221}$ & -0.08 &  F$_{4433}$ &  0.24 \\
F$_{222}$ &-24.1460 &  F$_{2222}$ &125.41 &  F$_{4444}$ &  0.61 \\                      
\hline
\end{tabular}


\end{table}
\endgroup

\begingroup
\begin{table*}

\caption{The CcCR $^{4}$HeHAr$^+$ Zero-Point ($R_{\alpha}$
vibrationally-averaged) and Equilibrium Structures, Rotation Constants,
Vibrationally Excited-Rotation Constants, Distortion Constants, and Vibrational
Frequencies (Intensities in km/mol).}

\label{4HeHAr}

\centering

\small

\begin{tabular}{l | l l l l l l}
\hline\hline

		   & HeH$^{36}$Ar$^+$ & HeD$^{36}$Ar$^+$ & HeH$^{38}$Ar$^+$ &
HeD$^{38}$Ar$^+$ & HeH$^{40}$Ar$^+$ & HeD$^{40}$Ar$^+$ \\

\hline
r$_0$(He$-$H) \AA  & 1.557 669 & 1.564 906 & 1.557 493 & 1.564 738 & 1.557 333 & 1.564 585 \\
r$_0$(Ar$-$H) \AA  & 1.275 998 & 1.282 695 & 1.276 025 & 1.282 729 & 1.276 048 & 1.282 759 \\
r$_e$(He$-$H) \AA  & 1.516 393 & 1.516 393 & 1.516 393 & 1.516 393 & 1.516 393 & 1.516 393 \\
r$_e$(Ar$-$H) \AA  & 1.294 727 & 1.294 727 & 1.294 727 & 1.294 727 & 1.294 727 & 1.294 727 \\
$B_e$ MHz          & 17147.8 & 16607.3 & 17044.4 & 16494.9 & 16951.2 & 16393.3 \\
$B_0$ MHz          & 16379.5 & 15900.3 & 16282.8 & 15794.8 & 16195.6 & 15699.4 \\
$B_1$ MHz          & 17288.3 & 16480.7 & 17184.8 & 16369.7 & 17091.5 & 16269.3 \\
$B_2$ MHz          & 15869.0 & 15757.4 & 15776.0 & 15644.1 & 15692.1 & 15542.1 \\
$B_3$ MHz          & 14955.0 & 14191.9 & 14871.1 & 14121.3 & 14795.4 & 14056.0 \\
$D_e$ kHz          & 434.842 & 373.927 & 429.277 & 368.279 & 424.287 & 363.322 \\
$H_e$ Hz           & -40.138 & -33.640 & -39.380 & -32.895 & -38.703 & -32.232 \\
$\mu$ D            & & & & & 1.48    & \\
\hline                                                    
$\omega_1(\sigma)$ Ar$-$H cm$^{-1}$ & 2542.8 & 1824.4 & 2540.9 & 1821.7 & 2539.2 (703) & 1819.3 \\
$\omega_2(\pi)$ Bend cm$^{-1}$      &  316.5 &  230.3 &  316.5 &  230.2 &  316.4 (158) &  230.1 \\
$\omega_3(\sigma)$ He$-$H cm$^{-1}$ &  220.1 &  219.3 &  219.6 &  219.2 &  219.1 (7)   &  219.7 \\
$\nu_1(\sigma)$ Ar$-$H cm$^{-1}$    & 2469.3 & 1791.3 & 2467.7 & 1788.9 & 2466.2       & 1786.8 \\
$\nu_2(\pi)$ Bend cm$^{-1}$         &  506.7 &  333.9 &  506.5 &  333.8 &  506.4       &  333.6 \\
$\nu_3(\sigma)$ He$-$H cm$^{-1}$    &  241.2 &  220.0 &  240.7 &  219.6 &  240.3       &  219.2 \\
$2\nu_1(\sigma)$ cm$^{-1}$          & 4724.5 & 3474.9 & 4724.5 & 3470.4 & 4721.7       & 3466.3 \\
$2\nu_2(\sigma)$ cm$^{-1}$          & 1120.2 &  726.1 & 1119.7 &  725.6 & 1119.4       &  725.2 \\
$2\nu_3(\sigma)$ cm$^{-1}$          &  431.2 &  389.0 &  430.5 &  388.4 &  429.8       &  387.9 \\
$\nu_1+\nu_2(\sigma)$ cm$^{-1}$     & 3087.4 & 2181.4 &  3085.7 & 2179.0 &       3084.2 & 2176.7 \\
$\nu_1+\nu_3(\sigma)$ cm$^{-1}$     &  2763.0 & 2048.2 & 2760.7 & 2045.3 &       2758.7 & 2042.7 \\
$\nu_2+\nu_3(\sigma)$ cm$^{-1}$     &  794.0 &  586.8 &  793.3 &  586.1 &        792.6 &  585.5 \\
ZPE cm$^{-1}$                       & 1809.4 & 1314.5 & 1808.1 & 1312.8 & 1806.9       & 1311.3 \\
\hline                                                    

\hline
\end{tabular}
\end{table*}

\endgroup

\renewcommand{\baselinestretch}{1}
\begingroup
\begin{table*}

\caption{The CcCR $^{3}$HeHAr$^+$ Zero-Point ($R_{\alpha}$
vibrationally-averaged) and Equilibrium Structures, Rotation Constants,
Vibrationally Excited-Rotation Constants, Distortion Constants, and Vibrational
Frequencies.}

\label{3HeHAr}

\centering

\small

\begin{tabular}{l | l l l l l l}
\hline\hline

		   & $^3$HeH$^{36}$Ar$^+$ & $^3$HeD$^{36}$Ar$^+$ &
$^3$HeH$^{38}$Ar$^+$ & $^3$HeD$^{38}$Ar$^+$ & $^3$HeH$^{40}$Ar$^+$ &
$^3$HeD$^{40}$Ar$^+$ \\

\hline
r$_0$(He$-$H) \AA  & 1.566 868 & 1.481 206 & 1.566 714 & 1.481 184 & 1.566 573 & 1.481 164 \\
r$_0$(Ar$-$H) \AA  & 1.276 728 & 1.272 902 & 1.276 754 & 1.272 956 & 1.276 777 & 1.273 004 \\
r$_e$(He$-$H) \AA  & 1.516 393 & 1.516 393 & 1.516 393 & 1.516 393 & 1.516 393 & 1.516 393 \\
r$_e$(Ar$-$H) \AA  & 1.294 727 & 1.294 727 & 1.294 727 & 1.294 727 & 1.294 727 & 1.294 727 \\
$B_e$ MHz          & 21849.5 & 20872.8 & 21742.7 & 20753.4 & 21646.2 & 20645.8 \\
$B_0$ MHz          & 20757.6 & 20949.4 & 20658.6 & 20829.3 & 20569.1 & 20720.9 \\
$B_1$ MHz          & 21789.1 & 20852.9 & 21773.0 & 20733.6 & 21677.1 & 20626.1 \\
$B_2$ MHz          & 20142.0 & 21016.9 & 20046.5 & 20896.0 & 19960.1 & 20787.0 \\
$B_3$ MHz          & 18683.6 & 21064.4 & 18600.2 & 20943.0 & 18524.7 & 20833.5 \\
$D_e$ kHz          & 682.267 & 554.180 & 675.061 & 546.953 & 668.595 & 540.474 \\
$H_e$ Hz           & -80.385 & -62.553 & -79.133 & -61.357 & -78.015 & -60.291 \\
\hline                                                    
$\omega_1(\sigma)$ Ar$-$H cm$^{-1}$ & 2542.9 & 1824.5 & 2541.0 & 1821.8 & 2539.2 & 1819.4 \\
$\omega_2(\pi)$ Bend cm$^{-1}$      &  319.1 &  233.8 &  319.1 &  233.8 &  319.0 &  233.7 \\
$\omega_3(\sigma)$ He$-$H cm$^{-1}$ &  250.5 &  233.8 &  250.0 &  233.8 &  249.6 &  233.7 \\
$\nu_1(\sigma)$ Ar$-$H cm$^{-1}$    & 2474.0 & 1803.8 & 2472.4 & 1801.5 & 2470.9 & 1799.4 \\
$\nu_2(\pi)$ Bend cm$^{-1}$         &  514.6 &  357.8 &  514.4 &  357.7 &  514.3 &  357.5 \\
$\nu_3(\sigma)$ He$-$H cm$^{-1}$    &  266.5 &  370.2 &  266.1 &  370.1 &  265.8 &  369.9 \\
$2\nu_1(\sigma)$ cm$^{-1}$          & 4737.0 & 3518.4 & 4734.0 & 3514.0 & 4731.3 & 3510.0 \\
$2\nu_2(\sigma)$ cm$^{-1}$          & 1138.5 &  780.6 & 1138.0 &  780.1 & 1137.7 &  779.7 \\
$2\nu_3(\sigma)$ cm$^{-1}$          &  466.7 &  791.6 &  466.1 &  791.2 &  465.6 &  790.9 \\
$\nu_1+\nu_2(\sigma)$ cm$^{-1}$     & 3101.0 & 2207.2 & 3099.3 & 2204.9 & 3097.8 & 2202.9 \\
$\nu_1+\nu_3(\sigma)$ cm$^{-1}$     & 2800.2 & 2219.6 & 2798.0 & 2217.3 & 2796.1 & 2215.3 \\
$\nu_2+\nu_3(\sigma)$ cm$^{-1}$     &  833.7 &  790.3 &  833.1 &  790.0 &  832.4 &  789.9 \\
ZPE cm$^{-1}$                       & 1831.9 & 1370.5 & 1830.6 & 1369.0 & 1829.4 & 1367.8 \\
\hline                                                    

\hline
\end{tabular}
\end{table*}

\endgroup

As the previously discussed energetics indicate, the behavior of HeHAr$^+$ is
quite different from the other two proton-bound complexes described in this
work and also from NeHNe$^+$.  The F$_{11}$ He$-$H force constant in HeHAr$^+$
is quite small at 0.105 mdyne/\AA$^2$ in Table \ref{HeHArfc}.  The F$_{22}$
Ar$-$H force constant is quite large at 4.011 mdyne/\AA$^2$, nearly the same
magnitude as that in argonium itself \citep{RTheis15}.  Table \ref{4HeHAr}
corroborates the strong Ar$-$H bonding and relatively weak He$-$H bonding in
the bond lengths themselves.  The 1.558 \AA\ He$-$H bond length is much longer
than any in the other two helium proton-bound complexes, and the 1.276 \AA\
Ar$-$H bond length is quite close to the 1.292 \AA\ bond length in argonium
\citep{Cueto14}.

The longer bond lengths and the heavier argon mass increase the rotational
constants such that they are roughly half as large as $B_v$ in HeHNe$^+$ which
are roughly half-again as large as $B_v$ in HeHHe$^+$.  The
center-of-mass-origin, CCSD/aug-cc-pVTZ dipole moment for HeH$^{40}$Ar$^+$ is
1.48 D smaller than that in HeHNe$^+$ making any rotational intensities less
for HeHAr$^+$ than in the neon complex.  The more weakly bonded nature of
HeHAr$^+$ is also carried out in increases of the quartic and sextic ($D_e$ and
$H_e$, respectively) distortion constants.  Both of these values are higher for
HeHAr$^+$ relative to HeHNe$^+$, and the -40.138 Hz $H_e$ in HeH$^{36}$Ar$^+$
is of greater magnitude than the same parameter in HeHHe$^+$.  The bond lengths
and the distortion constants, as well as the energetics, show that the
HeHAr$^+$ proton-bound complex is likely a strong van der Waals interaction
between the argonium cation and a helium atom and not a covalent interaction on
the helium end.

The intensities of the vibrational modes also belie a shift in the physical
construction of this complex.  The proton-shuttle motion is relatively dim
(although still absolutely bright) with an intensity of 703 km/mol.  The bend
and He$-$H stretch are also less intense, but they do not drop in value by as
much of a percentage relative to HeHHe$^+$ as $\omega_1$.  Similar behavior is
reported for ArHNe$^+$ \citep{Fortenberry17NgHNg+} making these trends highly
likely to be observed physically.

The $\nu_1$ proton shuttle motion now is actually more correctly described as
the Ar$-$H stretch since the He$-$H stretching component coefficient in
defining the potential shuttle motion normal mode is small.  The $\nu_1$
Ar$-$H stretch in HeH$^{36}$Ar$^+$ is 2469.3 cm$^{-1}$ (4.05 microns) which is
fairly close to the Ar$-$H stretch in argonium at 2612.5 cm$^{-1}$ (3.83
microns) \citep{Cueto14}.  The frequency of the $\nu_2$ bend and the $\nu_3$
He$-$H stretch drop significantly relative to the other complexes further
highlighting the relative weakness of the helium interactions with the
argonium core.  These two far-IR transitions are only weakly affected by the
change in argon mass, but, unexpectedly have a larger shift for inclusion of
the lighter $^{3}$He isotope as shown in Table \ref{3HeHAr}.  However, the
moderate intensity of the $\nu_2$ bend at 316.5 cm$^{-1}$ (31.60 microns)
makes this fundamental vibrational frequency a tantalizing candidate for
far-IR observation, especially for the next generation of space telescopes
currently within discussion of the presently ongoing decadal surveys.

\section{Conclusions}

The most striking result from this study in light of the earlier work by
\cite{Fortenberry17NgHNg+} is that neon and helium behave very similarly while
argon does not.  The intensities, fundamental vibrational frequencies, and
even binding energies are not significantly changed when moving down the
periodic table from helium to neon in such proton-bound complexes.  Once argon
is invoked, the chemistry changes fundamentally.  This is likely due to the
polarizability of argon and the energy proximity of the additional $d$
orbitals close to argon's valence orbital occupation.

In any case, the proton-bound complexes involving helium and neon (HeHHe$^+$,
HeHNe$^+$, and NeHNe$^+$) give very intense proton shuttle motions in the range
where the mid-IR becomes the far-IR.  The newest generation of space-based
telescopes like the upcoming James Webb Space Telescope (JWST) or even the
Stratospheric Observatory for Infrared Spectroscopy can potentially be utilized
to observe these vibrational frequencies.  The HeHNe$^+$ dipole moment is also
large making this proton-bound complex a candidate for rotational observation
with ground-based telescopes as has been the common practice for half a
century.  The HeHAr$^+$ complex appears to be readily formed from hypothesized
interstellar species, the helonium and argon trihydride cations.  HeHAr$^+$ is
also rotationally active and has a bright fundamental vibrational frequency,
but both of these are reduced relative to HeHNe$^+$.  

The sheer abundance of helium and hydrogen as well as the relatively high
abundance of neon make any molecules comprised of these species notable for
interstellar chemistry and subsequent astrophysical observation.  The relative
energetics, zero-point energies, and spectroscopic data provided in this work
will enhance further laboratory, modeling, or interstellar studies.  While
estimates of the actual abundances are beyond the scope of the present work,
the formation energetics imply that the abundances of these proton-bound
complexes are likely dependent upon the abundances of the possible precursors,
notably the noble gas-hydride and -trihydride cations.  As a result, HeHAr$^+$
will likely have the highest abundance.  However, the significant intensities
present for the proton shuttle fundamental frequencies of the helium and neon
complexes should make relatively small abundances for any of these noble gas
species detectable with JWST.  Consequently, the present work is showing a
likely case where more, natural noble gas molecules may be detected in the ISM
and the chemistries of these atoms can be further enhanced.

\section{Acknowledgements}

Georgia Southern University provided start-up funds in support of this work.
Additionally, Prof.~T.~Daniel Crawford of Virginia Tech provided some computer
hardware necessary to complete some of this work.  The authors are grateful for
the support from both.


\begin{thebibliography}{81}
\expandafter\ifx\csname natexlab\endcsname\relax\def\natexlab#1{#1}\fi

\bibitem[{Allen \& coworkers(2005)}]{intder}
Allen W.~D., coworkers, 2005. $INTDER\ 2005$ is a General Program Written by W.
  D. Allen and Coworkers, which Performs Vibrational Analysis and Higher-Order
  Non-Linear Transformations.

\bibitem[{Baccarelli, Gianturco \& Schneider(1997)Baccarelli, Gianturco, \&
  Schneider}]{Baccarelli97}
Baccarelli I., Gianturco F.~A., Schneider F., 1997, J. Phys. Chem. A, 101, 6054

\bibitem[{Barlow {et~al}\mbox{.}(2013)Barlow, Swinyard, Owen, Cernicharo,
  Gomez, Ivison, Krause, Lim, Matsuura, Miller, Olofsson, \&
  Polehampton}]{Barlow13}
Barlow M.~J. {et~al.}, 2013, Science, 342, 1343

\bibitem[{Bartl {et~al}\mbox{.}(2013)Bartl, Leidlmair, Denifl, Scheier, \&
  Echt}]{Bartl13}
Bartl P., Leidlmair C., Denifl S., Scheier P., Echt O., 2013, Chem. Phys.
  Chem., 14, 227

\bibitem[{Begum \& Subramanian(2016)}]{Begum16}
Begum S., Subramanian R., 2016, J. Molec. Model., 22, 6

\bibitem[{Borocci, Giordani \& Grandinetti(2015)Borocci, Giordani, \&
  Grandinetti}]{Borocci15}
Borocci S., Giordani M., Grandinetti F., 2015, J. Phys. Chem. A, 119, 6528

\bibitem[{Cotton {et~al}\mbox{.}(2012)Cotton, Francisco, Linguerri, \&
  Mitrushchenkov}]{Cotton12}
Cotton C.~E., Francisco J.~S., Linguerri R., Mitrushchenkov A.~O., 2012, J.
  Chem. Phys., 136, 184307

\bibitem[{Coxon \& Hajigeorgiou(2016)}]{Coxon16}
Coxon J.~A., Hajigeorgiou P.~G., 2016, J. Molec. Spectrosc., 330, 63

\bibitem[{Crawford \& {Schaefer III}(2000)}]{ccreview}
Crawford T.~D., {Schaefer III} H.~F., 2000, in Reviews in Computational
  Chemistry, Lipkowitz K.~B., Boyd D.~B., eds., Vol.~14, Wiley, New York, pp.
  33--136

\bibitem[{Cueto {et~al}\mbox{.}(2014)Cueto, Cernicharo, Barlow, Swinyard,
  Herrero, Tanarro, \& Dom\'{e}nech}]{Cueto14}
Cueto M., Cernicharo J., Barlow M.~J., Swinyard B.~M., Herrero V.~J., Tanarro
  I., Dom\'{e}nech J.~L., 2014, Astrophys. J., 78, L5

\bibitem[{Douglas \& Kroll(1974)}]{Douglas74}
Douglas M., Kroll N., 1974, Ann. Phys., 82, 89

\bibitem[{Dunning(1989)}]{Dunning89}
Dunning T.~H., 1989, J. Chem. Phys., 90, 1007

\bibitem[{Dykstra(1983)}]{Dykstra82}
Dykstra C.~E., 1983, J. Molec. Struct., 12, 131

\bibitem[{Fortenberry(2017)}]{Fortenberry17NgHNg+}
Fortenberry R.~C., 2017, ACS Earth Space Chem., 1, 60

\bibitem[{Fortenberry {et~al}\mbox{.}(2013)Fortenberry, Huang, Crawford, \&
  Lee}]{Fortenberry13C3H-}
Fortenberry R.~C., Huang X., Crawford T.~D., Lee T.~J., 2013, Astrophys. J.,
  772, 39

\bibitem[{Fortenberry {et~al}\mbox{.}(2014)Fortenberry, Huang, Crawford, \&
  Lee}]{Fortenberry14C2H3+}
Fortenberry R.~C., Huang X., Crawford T.~D., Lee T.~J., 2014, J. Phys. Chem. A,
  118, 7034

\bibitem[{Fortenberry {et~al}\mbox{.}(2011{\natexlab{a}})Fortenberry, Huang,
  Francisco, Crawford, \& Lee}]{Fortenberry11HOCO}
Fortenberry R.~C., Huang X., Francisco J.~S., Crawford T.~D., Lee T.~J.,
  2011{\natexlab{a}}, J. Chem. Phys., 135, 134301

\bibitem[{Fortenberry {et~al}\mbox{.}(2011{\natexlab{b}})Fortenberry, Huang,
  Francisco, Crawford, \& Lee}]{Fortenberry11cHOCO}
Fortenberry R.~C., Huang X., Francisco J.~S., Crawford T.~D., Lee T.~J.,
  2011{\natexlab{b}}, J. Chem. Phys., 135, 214303

\bibitem[{Fortenberry {et~al}\mbox{.}(2012)Fortenberry, Huang, Francisco,
  Crawford, \& Lee}]{Fortenberry12HOCO+}
Fortenberry R.~C., Huang X., Francisco J.~S., Crawford T.~D., Lee T.~J., 2012,
  J. Chem. Phys., 136, 234309

\bibitem[{Fortenberry, Lee \& Francisco(2016{\natexlab{a}})Fortenberry, Lee, \&
  Francisco}]{Fortenberry16COHNN+}
Fortenberry R.~C., Lee T.~J., Francisco J.~S., 2016{\natexlab{a}}, J. Phys.
  Chem A, 120, 7745

\bibitem[{Fortenberry, Lee \& Francisco(2016{\natexlab{b}})Fortenberry, Lee, \&
  Francisco}]{Fortenberry16NNHCO+}
Fortenberry R.~C., Lee T.~J., Francisco J.~S., 2016{\natexlab{b}}, Astrophys.
  J., 819, 141

\bibitem[{Fortenberry, Lee \& M\"{u}ller(2015)Fortenberry, Lee, \&
  M\"{u}ller}]{Fortenberry15SiC2}
Fortenberry R.~C., Lee T.~J., M\"{u}ller H. S.~P., 2015, Molec. Astrophys., 1,
  13

\bibitem[{Fortenberry, Roueff \& Lee(2016)Fortenberry, Roueff, \&
  Lee}]{Fortenberry16C2H3+}
Fortenberry R.~C., Roueff E., Lee T.~J., 2016, Chem. Phys. Lett., 650, 126

\bibitem[{Fortenberry {et~al}\mbox{.}(2015)Fortenberry, Yu, Mancini, Bowman,
  Lee, Crawford, Klemperer, \& Francisco}]{Fortenberry15OCHCO+}
Fortenberry R.~C., Yu Q., Mancini J.~S., Bowman J.~M., Lee T.~J., Crawford
  T.~D., Klemperer W.~F., Francisco J.~S., 2015, J. Chem. Phys., 143, 071102

\bibitem[{Frenking \& Cremer(1990)}]{Frenking90}
Frenking G., Cremer D., 1990, Struct. Bond., 73, 17

\bibitem[{Fridgen \& Parnis(1998)}]{Fridgen98}
Fridgen T.~D., Parnis J.~M., 1998, J. Chem. Phys., 109, 2162

\bibitem[{Frisch {et~al}\mbox{.}(2009)Frisch, Trucks, Schlegel, Scuseria, Robb,
  Cheeseman, Scalmani, Barone, Mennucci, Petersson, Nakatsuji, Caricato, Li,
  Hratchian, Izmaylov, Bloino, Zheng, Sonnenberg, Hada, Ehara, Toyota, Fukuda,
  Hasegawa, Ishida, Nakajima, Honda, Kitao, Nakai, Vreven, Montgomery, Peralta,
  Ogliaro, Bearpark, Heyd, Brothers, Kudin, Staroverov, Kobayashi, Normand,
  Raghavachari, Rendell, Burant, Iyengar, Tomasi, Cossi, Rega, Millam, Klene,
  Knox, Cross, Bakken, Adamo, Jaramillo, Gomperts, Stratmann, Yazyev, Austin,
  Cammi, Pomelli, Ochterski, Martin, Morokuma, Zakrzewski, Voth, Salvador,
  Dannenberg, Dapprich, Daniels, Farkas, Foresman, Ortiz, Cioslowski, \&
  Fox}]{g09}
Frisch M.~J. {et~al.}, 2009, Gaussian~09 {R}evision {D}.01. Gaussian Inc.
  Wallingford CT 

\bibitem[{Gamallo, Huarte-Larranaga \& Gonz\'{a}lez(2013)Gamallo,
  Huarte-Larranaga, \& Gonz\'{a}lez}]{Gamallo13}
Gamallo P., Huarte-Larranaga F., Gonz\'{a}lez M., 2013, J. Phys. Chem. A, 117,
  5393

\bibitem[{Gaw {et~al}\mbox{.}(1991)Gaw, Willets, Green, \& Handy}]{spectro91}
Gaw J.~F., Willets A., Green W.~H., Handy N.~C., 1991, in Advances in Molecular
  Vibrations and Collision Dynamics, Bowman J.~M., Ratner M.~A., eds., JAI
  Press, Inc., Greenwich, Connecticut, pp. 170--185

\bibitem[{Grabowski {et~al}\mbox{.}(2016)Grabowski, Ugalde, Andrada, \&
  Frenking}]{Grabowski16}
Grabowski S.~J., Ugalde J.~M., Andrada D.~M., Frenking G., 2016, Chem. Euro.
  J., 22, 11317

\bibitem[{Grandinetti(2004)}]{Grandinetti04}
Grandinetti F., 2004, Int. J. Mass Spectrom., 237, 243

\bibitem[{Grandinetti(2011)}]{Grandinetti11}
Grandinetti F., 2011, Eur. J. Mass Spectrom., 17, 423

\bibitem[{Hehre, Ditchfeld \& Pople(1972)Hehre, Ditchfeld, \& Pople}]{Hehre72}
Hehre W.~J., Ditchfeld R., Pople J.~A., 1972, J. Chem. Phys., 56, 2257

\bibitem[{Hogness \& Lunn(1925)}]{Hogness25}
Hogness T.~R., Lunn E.~G., 1925, Phys. Rev., 26, 44

\bibitem[{Huang, Fortenberry \& Lee(2013{\natexlab{a}})Huang, Fortenberry, \&
  Lee}]{Huang13NNOH+}
Huang X., Fortenberry R.~C., Lee T.~J., 2013{\natexlab{a}}, J. Chem. Phys.,
  139, 084313

\bibitem[{Huang, Fortenberry \& Lee(2013{\natexlab{b}})Huang, Fortenberry, \&
  Lee}]{Huang13C3H+}
Huang X., Fortenberry R.~C., Lee T.~J., 2013{\natexlab{b}}, Astrophys. J.
  Lett., 768, 25

\bibitem[{Huang \& Lee(2008)}]{Huang08}
Huang X., Lee T.~J., 2008, J. Chem. Phys., 129, 044312

\bibitem[{Huang \& Lee(2009)}]{Huang09}
Huang X., Lee T.~J., 2009, J. Chem. Phys., 131, 104301

\bibitem[{Huang, Taylor \& Lee(2011)Huang, Taylor, \& Lee}]{Huang11}
Huang X., Taylor P.~R., Lee T.~J., 2011, J. Phys. Chem. A, 115, 5005

\bibitem[{Kendall, Dunning \& Harrison(1992)Kendall, Dunning, \&
  Harrison}]{aug-cc-pVXZ}
Kendall R.~A., Dunning T.~H., Harrison R.~J., 1992, J. Chem. Phys., 96, 6796

\bibitem[{Kitchens \& Fortenberry(2016)}]{Kitchens16}
Kitchens M. J.~R., Fortenberry R.~C., 2016, Chem. Phys., 472, 119

\bibitem[{Koner {et~al}\mbox{.}(2014)Koner, Barrios, Gonzalez-Lezana, \&
  Panda}]{Koner14}
Koner D., Barrios L., Gonzalez-Lezana T., Panda A.~N., 2014, J. Chem. Phys.,
  141, 114302

\bibitem[{Koner {et~al}\mbox{.}(2016)Koner, Barrios, Gonzalez-Lezana, \&
  Panda}]{Koner16}
Koner D., Barrios L., Gonzalez-Lezana T., Panda A.~N., 2016, J. Chem. Phys.,
  144, 034303

\bibitem[{Koner, Vats \& Panda(2012)Koner, Vats, \& Panda}]{Koner12}
Koner D., Vats A., Panda M. V. A.~N., 2012, Comput. and Theoret. Chem., 1000,
  19

\bibitem[{Lundell, Pettersson \& Rasanen(1999)Lundell, Pettersson, \&
  Rasanen}]{Lundell99}
Lundell J., Pettersson M., Rasanen M., 1999, Phys. Chem. Chem. Phys., 1, 181

\bibitem[{Martin \& Lee(1996)}]{Martin96}
Martin J. M.~L., Lee T.~J., 1996, Chem. Phys. Lett., 258, 136

\bibitem[{Martin \& Taylor(1994)}]{Martin94}
Martin J. M.~L., Taylor P.~R., 1994, Chem. Phys. Lett., 225, 473

\bibitem[{Matsushima {et~al}\mbox{.}(1998)Matsushima, Ohtaki, Torige, \&
  Takagi}]{Matsashima98}
Matsushima F., Ohtaki Y., Torige O., Takagi K., 1998, J. Chem. Phys., 109, 2242

\bibitem[{{McDonald II} {et~al}\mbox{.}(2016){McDonald II}, Mauney, Leicht,
  Marks, Tan, Kuo, \& Duncan}]{McDonald16}
{McDonald II} D.~C., Mauney D.~T., Leicht D., Marks J.~H., Tan J.~A., Kuo
  J.-L., Duncan M.~A., 2016, J. Chem. Phys., 145, 231101

\bibitem[{Mills(1972)}]{Mills72}
Mills I.~M., 1972, in Molecular Spectroscopy - Modern Research, Rao K.~N.,
  Mathews C.~W., eds., Academic Press, New York, pp. 115--140

\bibitem[{M{\o}ller \& Plesset(1934)}]{MP2}
M{\o}ller C., Plesset M.~S., 1934, Phys. Rev., 46, 618

\bibitem[{Neufeld \& Wolfire(2016)}]{Neufeld16}
Neufeld D.~A., Wolfire M.~G., 2016, Astrophys. J., 826, 183

\bibitem[{Novak \& Fortenberry(2017)}]{Novak17}
Novak C.~M., Fortenberry R.~C., 2017, Phys. Chem. Chem. Phys., 19, 5230

\bibitem[{Panda \& Sathyamurthy(2003)}]{Panda03}
Panda A.~N., Sathyamurthy N., 2003, J. Phys. Chem. A, 107, 7125

\bibitem[{Papousek \& Aliev(1982)}]{Papousek82}
Papousek D., Aliev M.~R., 1982, Molecular Vibration-Rotation Spectra. Elsevier,
  Amsterdam

\bibitem[{Pauzat \& Ellinger(2005)}]{Pauzat05}
Pauzat F., Ellinger Y., 2005, Planet. Space Sci., 53, 1389

\bibitem[{Pauzat \& Ellinger(2007)}]{Pauzat07}
Pauzat F., Ellinger Y., 2007, J. Chem. Phys., 127, 014308

\bibitem[{Pauzat {et~al}\mbox{.}(2013)Pauzat, Ellinger, Mousis, Dib, \&
  Ozgurel}]{Pauzat13}
Pauzat F., Ellinger Y., Mousis O., Dib M.~A., Ozgurel O., 2013, Astrophys. J.,
  777, 29

\bibitem[{Pauzat {et~al}\mbox{.}(2009)Pauzat, Ellinger, Pilm\`{e}, \&
  Mousis}]{Pauzat09}
Pauzat F., Ellinger Y., Pilm\`{e} J., Mousis O., 2009, J. Chem. Phys., 130,
  174313

\bibitem[{Peterson \& Dunning(1995)}]{cc-pVXZ}
Peterson K.~A., Dunning T.~H., 1995, J. Chem. Phys., 102, 2032

\bibitem[{Raghavachari {et~al}\mbox{.}(1989)Raghavachari, Trucks., Pople, \&
  Head-Gordon}]{Rag89}
Raghavachari K., Trucks. G.~W., Pople J.~A., Head-Gordon M., 1989, Chem. Phys.
  Lett., 157, 479

\bibitem[{Ram, Bernath \& Brault(1985)Ram, Bernath, \& Brault}]{Ram85}
Ram R., Bernath P., Brault J., 1985, J. Molec. Spectrosc., 113, 451

\bibitem[{Rice {et~al}\mbox{.}(1991)Rice, Taylor, Lee, \& Alml\"{o}f}]{Rice91}
Rice J.~E., Taylor P.~R., Lee T.~J., Alml\"{o}f J., 1991, J. Chem. Phys., 94,
  4972

\bibitem[{Roberge \& Dalgarno(1982)}]{Roberge82}
Roberge W., Dalgarno A., 1982, Astrophys. J., 255, 489

\bibitem[{Roth, Dopfer \& Maier(2001)Roth, Dopfer, \& Maier}]{Roth01}
Roth D., Dopfer O., Maier J.~P., 2001, Phys. Chem. Chem. Phys., 3, 2400

\bibitem[{Roueff, Alekseyev \& Bourlot(2014)Roueff, Alekseyev, \&
  Bourlot}]{Roueff14}
Roueff E., Alekseyev A.~B., Bourlot J.~L., 2014, Astron. Astrophys., 566, A30

\bibitem[{Savage \& Sembach(1996)}]{Savage96}
Savage B.~D., Sembach K.~R., 1996, Annu. Rev. Astron. Astrophys., 34, 279

\bibitem[{Savic {et~al}\mbox{.}(2015)Savic, Gerlich, Asvany, Jusko, \&
  Schlemmer}]{Savic15}
Savic I., Gerlich D., Asvany O., Jusko P., Schlemmer S., 2015, Mol. Phys., 113,
  2320

\bibitem[{Schilke {et~al}\mbox{.}(2014)Schilke, Neufeld, M\"{u}ller, Comito,
  Bergin, Lis, Gerin, Black, Wolfire, Indriolo, Pearson, Menten, Winkel,
  S\'{a}nchez-Monge, M\"{o}ller, Godard, \& Falgarone}]{Schilke14}
Schilke P. {et~al.}, 2014, Astron. Astrophys., 566, A29

\bibitem[{Shavitt \& Bartlett(2009)}]{Shavitt09}
Shavitt I., Bartlett R.~J., 2009, Many-Body Methods in Chemistry and Physics:
  MBPT and Coupled-Cluster Theory. Cambridge University Press, Cambridge

\bibitem[{Taylor {et~al}\mbox{.}(1989)Taylor, Lee, Rice, \&
  Alml\"{o}f}]{Taylor89}
Taylor P.~R., Lee T.~J., Rice J.~E., Alml\"{o}f J., 1989, Chem. Phys. Lett.,
  163, 359

\bibitem[{Terrill \& Nesbitt(2010)}]{Terrill10}
Terrill K., Nesbitt D.~J., 2010, Phys. Chem. Chem. Phys., 12, 8311

\bibitem[{Theis \& Fortenberry(2015)}]{RTheis152}
Theis R.~A., Fortenberry R.~C., 2015, J. Phys. Chem. A, 119, 4915

\bibitem[{Theis \& Fortenberry(2016)}]{RTheis16}
Theis R.~A., Fortenberry R.~C., 2016, Mol. Astrophys., 2, 18

\bibitem[{Theis, Morgan \& Fortenberry(2015)Theis, Morgan, \&
  Fortenberry}]{RTheis15}
Theis R.~A., Morgan W.~J., Fortenberry R.~C., 2015, Mon. Not. R. Astron. Soc.,
  446, 195

\bibitem[{Turney {et~al}\mbox{.}(2012)Turney, Simmonett, Parrish, Hohenstein,
  Evangelista, Fermann, Mintz, Burns, Wilke, Abrams, Russ, Leininger, Janssen,
  Seidl, Allen, {Schaefer III}, King, Valeev, Sherrill, \& Crawford}]{psi4}
Turney J.~M. {et~al.}, 2012, Wiley Interdisciplinary Reviews: Computational
  Molecular Science, 2, 556

\bibitem[{Watson(1977)}]{Watson77}
Watson J. K.~G., 1977, in Vibrational Spectra and Structure, During J.~R., ed.,
  Elsevier, Amsterdam, pp. 1--89

\bibitem[{Werner {et~al}\mbox{.}(2010)Werner, Knowles, Manby, {Sch\"{u}tz},
  Celani, Knizia, Korona, Lindh, Mitrushenkov, Rauhut, Adler, Amos,
  Bernhardsson, Berning, Cooper, Deegan, Dobbyn, Eckert, Goll, Hampel,
  Hesselmann, Hetzer, Hrenar, Jansen, K\"oppl, Liu, Lloyd, Mata, May,
  McNicholas, Meyer, Mura, Nicklass, Palmieri, Pfl\"uger, Pitzer, Reiher,
  Shiozaki, Stoll, Stone, Tarroni, Thorsteinsson, Wang, \& Wolf}]{molpro}
Werner H.-J. {et~al.}, 2010, Molpro, version 2010.1, a package of $ab\ initio$
  programs. See http://www.molpro.net

\bibitem[{Yu {et~al}\mbox{.}(2015)Yu, Bowman, Fortenberry, Mancini, Lee,
  Crawford, Klemperer, \& Francisco}]{Yu15NNHNN+}
Yu Q., Bowman J.~M., Fortenberry R.~C., Mancini J.~S., Lee T.~J., Crawford
  T.~D., Klemperer W., Francisco J.~S., 2015, J. Phys. Chem. A, 119, 11623

\bibitem[{Zhao, Doney \& Linnartz(2014)Zhao, Doney, \& Linnartz}]{Zhao14}
Zhao D., Doney K.~D., Linnartz H., 2014, Astrophys. J. Lett., 791, L28

\bibitem[{Zicler {et~al}\mbox{.}(2016)Zicler, {Bacchus-Montabonel}, Pauzat,
  Chaquin, \& Ellinger}]{Zicler16}
Zicler E., {Bacchus-Montabonel} M.-C., Pauzat F., Chaquin P., Ellinger Y.,
  2016, J. Chem. Phys., 144, 111103

\end{thebibliography}



\end{document}